\documentclass[pra,reprint,floatfix,showpacs]{revtex4-1}

\usepackage{amsmath,amsfonts,amssymb}
\usepackage{graphicx}

\def\br {{\bf r}}

\begin{document}

\title{Breaking the resilience of a two-dimensional Bose-Einstein condensate to fragmentation}

\author{Shachar Klaiman$^{1,*}$\footnote[0]{$^*$ 
Corresponding author: Shachar.Klaiman@pci.uni-heidelberg.de},
Axel U. J. Lode$^{2}$, Alexej I. Streltsov$^1$,\\ 
Lorenz S. Cederbaum$^1$, and Ofir E. Alon$^3$}

\affiliation{$^1$ Theoretische Chemie, Physikalisch--Chemisches Institut, Universit\"at Heidelberg, 
Im Neuenheimer Feld 229, D-69120 Heidelberg, Germany}
\affiliation{$^2$ Department of Physics, University of Basel, Klingelbergstrasse 82, CH-4056 Basel, Switzerland}
\affiliation{$^3$ Department of Physics, University of Haifa at Oranim, Tivon 36006, Israel}

\date{\today}

\begin{abstract}
A two-dimensional Bose-Einstein condensate (BEC) 
split by a radial potential barrier is investigated.
We determine on an accurate many-body level the system's ground-state phase diagram 
as well as a time-dependent phase diagram of the splitting process.
Whereas the ground state is condensed for a wide range of parameters,
the time-dependent splitting process leads to substantial fragmentation.
We demonstrate for the first time the dynamical fragmentation of a BEC 
despite its ground state being condensed.
The results are analyzed by a mean-field model and suggest 
that a large manifold of low-lying fragmented excited states 
can significantly impact the dynamics of trapped two-dimensional BECs.
\end{abstract}

\pacs{03.75.Kk, 05.30.Jp, 03.65.-w}

\maketitle 

Shortly after the first experimental demonstration of 
trapped Bose-Einstein condensates (BECs) 
in three dimensions \cite{BEC1,BEC2,BEC3},
BECs in two-dimensional traps have been realized \cite{2D_BEC_K,2D_BEC_G,RMP3}.
While three-dimensional trapped BECs have been extensively studied since their discovery, 
the static and time-dependent properties of their
two-dimensional counterparts are comparatively less explored. 

One of the most popular scenarios studied with ultracold bosonic atoms,
both experimentally and theoretically, 
is the splitting of a BEC by a central barrier into two spatially-disjoint clouds, e.g., 
Refs.~\cite{sp1,sp2,sp3,sp4,sp5,sp6,sp7,sp8,sp9,sp10,sp11,MCTDHB1,MCTDHB_OCT,BR,Muga}.
It is a common practice that in order to produce a fragmented BEC in the splitting process,
the ground state must be fragmented. 
This renders high barriers and strong interaction strengths necessary.
Previous works dealt with splitting of a BEC in one or three spatial dimensions.
To the best of our knowledge, splitting of a BEC in two spatial dimensions  
has not been explored experimentally or theoretically on the many-body level.

\begin{figure}[!]
\includegraphics[width=\columnwidth,angle=0]{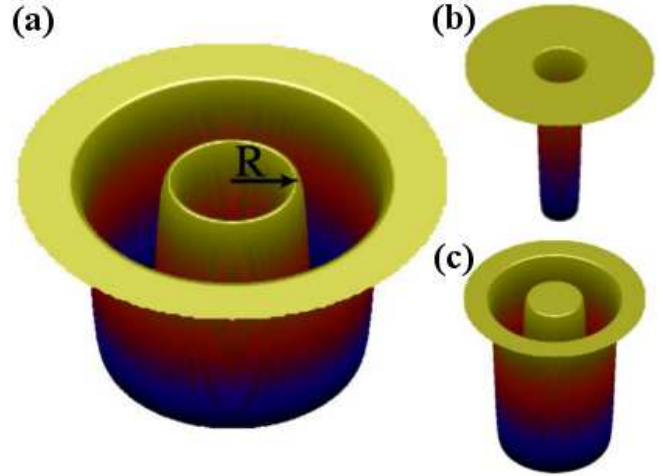}
\caption{(Color online) 
(a) The two-dimensional circular trap split by a radial barrier of radius $R$. 
(b) The inner disk and (c) outer annulus model potentials 
used to interpret the ground-state phase diagram.}
\label{f1}
\end{figure}

In the present work we investigate theoretically, 
on an accurate many-body level,
the physics of splitting a two-dimensional (2D) BEC.
A natural approach is to exploit the 2D symmetry of the system.
We thus split a circular BEC by a radial potential barrier,
see Fig.~\ref{f1}a for an illustration. 
This would lead to two concentric clouds,
unlike the above-discussed common 
way of splitting a BEC and,
as we shall see below, 
enrich the physics of BEC splitting.

By analyzing the many-body time-independent and time-dependent wavefunctions of the system, 
we construct both static and dynamic phase diagrams of the splitting process.
Whereas the ground state is condensed for a wide range of parameters, 
the time-dependent splitting process leads to substantial fragmentation.
We therefore demonstrate the dynamical fragmentation of a BEC, 
{\it despite} its ground state being fully condensed.
The results imply that a large manifold of fragmented excited 
states can significantly impact the dynamics of 2D BECs. 

We consider a repulsive BEC with $N=100$ bosons
in the 2D circular trap shown in Fig.~\ref{f1}a.
Throughout this work dimensionless units are used, 
such that the single-particle kinetic-energy operator reads
$\hat T(\br) = -\frac{1}{2} \nabla_{\br}^2$ \cite{units1}.
The explicit form of the one-body potential is given by
$V(\br) = V_{trap}(\br) + V_{barrier}(\br)$.
Here $V_{trap}(\br) = \{ 200 e^{-(r-r_c)^4/2}, r \le r_c = 9; 200, r>r_c\}$
is a flat trap which has the shape of ``a crater'' and
$V_{barrier}(\br) = 200 e^{-2(r-R)^4}$ 
a ringed-shaped radial barrier of radius $R$. 
We have chosen a flat potential $V_{trap}(\br)$ 
in order to allow the BEC to fill in the full area. 

It is instructive to commence with an analysis of the ground state of 
the {\it non-interacting} system.
The single-particle Schr\"odinger equation 
reads $\left[\hat T(\br) + V(\br)\right] f(\br) = \varepsilon f(\br)$.
The potential $V(\br)$
can be considered as made of two separated parts:
An inner disk and an outer annulus (see Fig.~\ref{f1}b,c),
separated by a radial barrier centered at $r=R$.
Obviously, the energy $\varepsilon$ of the particle changes with the barrier's radius $R$.
The energy of a particle in a disk of radius $R_d$ is well known and given by
$\varepsilon_{disk} = \frac{j_0^2}{2 R_d^2}$, where $j_0=2.4048$ is the first
zero of the zeroth Bessel function, e.g., Ref.~\cite{SIAM_1984}.
For an annulus of radii $R_{a1} < R_{a2}$, 
a remarkably precise (for not too small radii ratios) 
closed-form expression has recently been given in \cite{JMP_2011} and reads 
$\varepsilon_{annulus} \approx \frac{\ln^2(R_{a1}/R_{a2})+\pi^2}
{(R_{a1}^2 - R_{a2}^2)\ln(R_{a1}/R_{a2})}$.
These expressions allow us to determine, 
as a function of $R$,
where in the trap $V(\br)$ the particle is located.
For a high barrier, 
$\varepsilon_{disk} < \varepsilon_{annulus}$ implies
that the particle is located in the inner disk whereas the 
inverse relation $\varepsilon_{annulus} < \varepsilon_{disk}$ 
implies that it is localized in the outer annulus.
Beyond the obvious effect of the size of each part of the trap dictated by the radius $R$, 
in 2D one must also consider the naturally occurring attractive term originating from the kinetic energy. 
Since the ground state is radially symmetric, $f(\br)=f(r)$, 
and making the standard change of variables $f(r) \rightarrow \frac{f(r)}{\sqrt{r}}$,
one finds $\left[-\frac{1}{2}\frac{\partial^2}{\partial r^2} + V(r) - \frac{1}{8} \frac{1}{r^2}\right]f(r) 
= \varepsilon f(r)$.
Thus, for the ground state there is an effective attractive potential,
$V_{2D}(r)=-\frac{1}{8} \frac{1}{r^2}$,
pulling the particle towards the center.
This attractive force plays a crucial role in the physics described below.
Furthermore, one might expect that there is a critical $R$ 
for which $\varepsilon_{disk} = \varepsilon_{annulus}$
and the particle is located both in the disk and the annulus parts of the trap.
We will return to these points when the interaction is turned on,
and offer a generalization thereof. 

We now switch on the interaction between the particles and move 
to investigate its effect on the ground state of the system.
Specifically, we would like to study the one-body coherence properties 
of the ground state and ascertain when the many-body state 
is fragmented \cite{nozieres:82,nozieres:96,Spekkens,MCHB,Erich,fg1,fg2,fg3,fg4,fg5,fg6} 
or condensed \cite{Penrose}. 
This many-body property, 
which is derived from the eigenvalues of the reduced one-body 
density matrix \cite{Lowdin,RDMbook}, 
unambiguously conveys whether the BEC can be described within a single-orbital mean-field theory, 
i.e., the Gross-Pitaevski (GP) equation, 
or is it necessary to solve for the many-body state which spreads the bosons over many orbitals. 
Clearly, one can know this only \textit{a posteriori},
so we must solve the complicated 
many-body Hamiltonian in order to know 
whether the GP equation would have sufficed.

\begin{figure}[!]
\includegraphics[width=\columnwidth,angle=0]{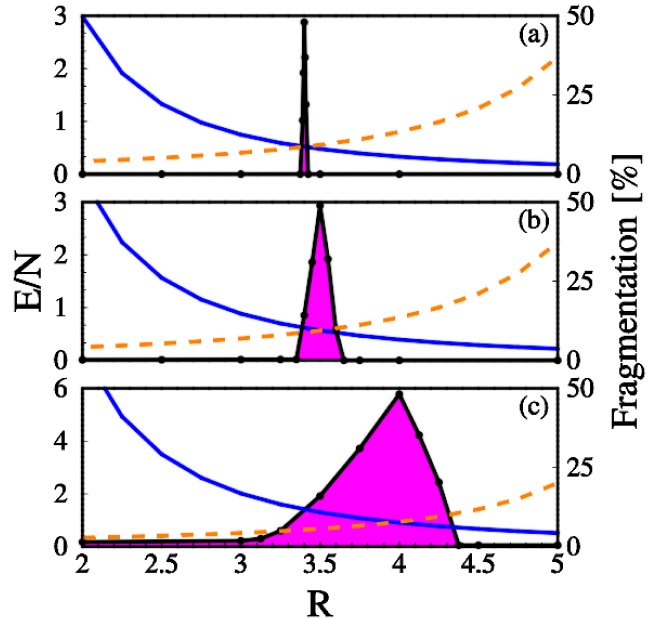}
\caption{(Color online) 
Ground-state many-body phase diagram of a BEC in a circular trap.
The three panels correspond from top to bottom to the interaction strengths 
$\lambda_0=0.002$, $0.02$, and $0.2$, respectively. 
The number of bosons is $N=100$.
The splitting changes the coherence properties of the BEC.
The BEC is mostly condensed, 
except for a narrow window (shaded magenta area) 
of radii $R$ which depends on $\lambda_0$.
The weaker the interaction is the narrower the window in which 
the BEC is fragmented.
A static model based on the GP theory is shown.
The energies $\varepsilon_{disk}^{GP}$ (solid blue) 
and $\varepsilon_{annulus}^{GP}$ (dashed orange) as a function of $R$ are depicted.
The maximal fragmentation on the many-body level is encountered when
$\varepsilon_{disk}^{GP} = \varepsilon_{annulus}^{GP}$.
All quantities are dimensionless.
\label{f2}}
\end{figure}

A suitable platform to study the many-body time evolution of trapped BECs is provided by the 
multiconfigurational time-dependent Hartree for bosons (MCTDHB) method \cite{MCTDHB1,MCTDHB2}. 
The MCTDHB method has been shown to produce accurate many-body solutions 
in various applications \cite{BJJ,MCTDHB_OCT,Benchmarks,LC_NJP,MCTDHB_2D_3D,MCTDHB_2D_3D_dyn}, 
and is well documented in the literature \cite{book1,book2}.
Until recently, MCTDHB has been applied to one-dimensional systems.
Most recently, MCTDHB has been implemented in higher dimensions \cite{MCTDHB_2D_3D_dyn},
which allows us now to enlarge the range of applications to 2D and three dimensions.
We use the implementation in the recursive MCTDHB (R-MCTDHB) \cite{R_MCTDHB}
and MCTDHB \cite{MCTDHB_Package} software packages. 

The many-boson Hamiltonian is given by
$\hat H(\br_1,\ldots,\br_N) = \sum_{j=1}^N [\hat T(\br_j) + V(\br_j)] + \sum_{j<k} W(\br_j-\br_k)$. 
The short-range repulsive interaction between the bosons is modeled by a Gaussian 
function \cite{Gauss,222}
$W(\br-\br') = \lambda_0 \frac{e^{-(\br-\br')^2/2\sigma^2}}{2\pi\sigma^2}$
with a width $\sigma=0.25$. 
The interaction parameter $\lambda_0$ is taken to be positive to describe repulsive bosons. 
A square box of size $[-12,12) \times [-12,12)$ and
spatial grid of size $128 \times 128$ were found to converge the results
to the accuracy given below.
In order to quantify the fragmentation of the many-body state, 
it is convenient to define 
it as the sum of all but the first eigenvalue of the reduced one-body density matrix. 

Figure~\ref{f2} depicts the ground-state fragmentation versus the position of the radial barrier 
for three different interaction strengths, $\lambda_0 = 0.002, 0.02, 0.2$ \cite{units2}. 
These many-body phase diagrams show that 
the radii $R$ for which the ground state is fragmented are very limited, 
namely that the ground state is mostly condensed within the parameter space of the problem. 
Increasing the interaction strength leads to two distinct effects. 
First, the maximal fragmentation 
shifts to larger values of $R$ and, second, 
the width of the fragmented region also increases with the interaction.
Importantly, we note that essentially $50\%$ fragmentation for different interaction strengths 
has been reached. 
The maxima occur for traps of different radii. 
Throughout this work we have preformed all computations with $4$
orbitals, and found that no more than two orbitals are macroscopically occupied.
Hence, the fragmentation of the BEC essentially equals
to the second eigenvalue of the reduced one-body density matrix.

\begin{figure}[!]
\includegraphics[width=1.00\columnwidth,angle=0]{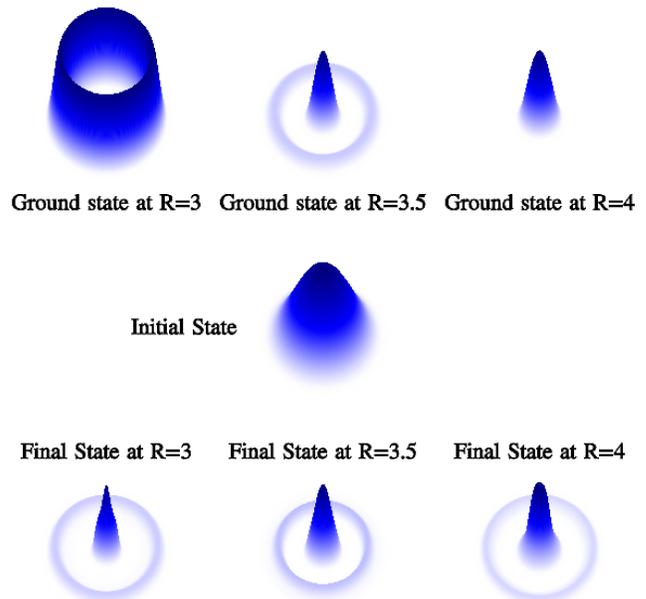}
\caption{(Color online)  The density of the ground state and of the 
wavefunction after the splitting process for $\lambda_0=0.02$.
Top row: For different values of $R$ the ground state is located either inside the inner disk part, 
outside in the annular part, or in a combination of both. 
Middle: Without the radial barrier, 
the BEC is spread out in the circular trap $V_{trap}(\br)$.
Bottom row: The density after the splitting process
is located both in the disk and annulus for a wide
range of radii $R$.}
\label{f3}
\end{figure}

In order to understand the phase diagrams depicted in Fig.~\ref{f2}, 
we set up a model. 
The model is based on the GP mean-field solutions of the $N$ interacting bosons 
in the inner disk and in the outer annulus parts
(see Fig.~\ref{f1}b,c). 
The GP energies per particle $\varepsilon_{disk}^{GP}$ and $\varepsilon_{annulus}^{GP}$ 
are depicted as a function of the barrier's position $R$ 
in Fig.~\ref{f2} for each interaction strength.
Remarkably, the intersection points of the two curves, 
$\varepsilon^{GP}_{disk}=\varepsilon^{GP}_{annulus}$, 
which mark a mean-field degeneracy between the inner and outer parts of the trap, 
accurately indicate the maximal fragmentation of the system on the many-body level. 
Moreover, the density of the fragmented ground state of the split BEC
occupies both the inner and the outer parts of the trap,
see Fig.~\ref{f3}. 
In the limit of weak interaction, 
our mean-field model connects with the non-interacting system discussed above. 
Namely, for the parameters of the potential studied, 
the non-interacting model predicts a degeneracy around $R=3.3$
which is in quite a good agreement with the maximal fragmentation 
in the case of the weak interaction, i.e., $R=3.4$. 

The results in Figs.~\ref{f2} and \ref{f3} 
clearly indicate that fragmentation of the BEC is accompanied by 
the spatial occupation of both the inner disk and the outer annulus of the trap. 
When the interaction energy is larger than the energy difference between
the disk and the annulus,
the BEC spreads over the two parts.
Consequently, the fragmented region in the phase diagram increases as the interaction becomes stronger. 
Within the fragmented region in the phase diagrams, 
the energy of the fragmented system is lower 
than the energy of the condensed system. 

Another interesting property of the phase diagrams is that the radius $R$ where the fragmentation 
is maximal increases with $\lambda_0$. 
For the disk and annular regions to be energetically equivalent (in the GP sense), 
the disk part should be smaller because of the attraction $V_{2D}(r)$ towards the center.
Hence, the GP orbital in the disk is more localized
than the GP orbital in the annulus.
Since the interaction energy scales like the fourth power of the GP orbital, 
when $\lambda_0$ is enlarged $R$ must increase 
in order to compensate for the growing interaction energy.

So far we have explored the static properties of the ground state showing it is mostly condensed.
One might expect that also dynamically splitting a BEC by raising a radial barrier would lead to a condensed state,
at the very least in the adiabatic limit when the radial barrier is raised slowly enough. 
It turns out that the dynamical picture is much more intriguing. 

To explore the dynamical process of splitting the BEC, 
we prepare the BEC in the ground state of the trap $V_{trap}(\br)$. 
In the absence of the radial barrier, 
the BEC is spread in the flat circular trap, see Fig~\ref{f3}.
One then ramps up the radial barrier such that the time-dependent one-body potential 
reads $V(\br,t) = V_{trap}(\br) + V_{ramp-up}(\br,t)$, 
where $V_{ramp-up}(\br,t) = \frac{\beta t}{200}V_{barrier}(\br)$ and $\beta$ 
is the splitting rate (the ramp-up process stops when the barrier 
reaches its maximal height, i.e., $\beta t = 200$).
This is a demanding many-body problem in 2D, 
because the BEC changes significantly 
both its shape {\it and} coherence, 
which MCTDHB can efficiently handle \cite{R_MCTDHB,MCTDHB_Package}.

\begin{figure}[!]
\includegraphics[width=1.00\columnwidth,angle=0]{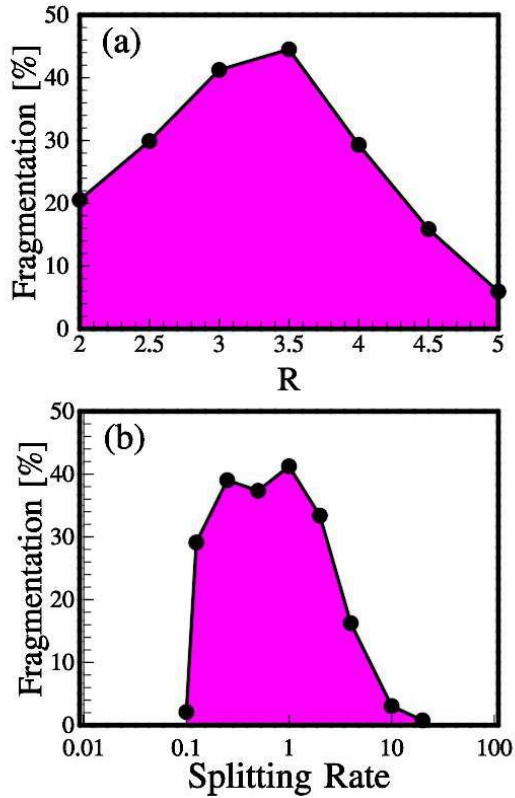}
\caption{(Color online) Two cuts through the time-dependent phase diagram 
for $N=100$ bosons and interaction strength $\lambda_0=0.02$. 
The fragmentation depicted refers to the end of the splitting process. 
(a) Remarkably, the dynamical splitting process leads to fragmentation 
over the entire examined range of radii $R$, 
see for comparison Fig.~\ref{f2}b.
The splitting rate is $\beta=1$. 
(b) The dynamical splitting process leads to fragmentation 
over two orders of magnitude of the splitting rate.
The radius is $R=3$. 
In the splitting process the system has a high affinity 
to fragment.
All quantities are dimensionless.
\label{f4}}
\end{figure}

Fig.~\ref{f4}a depicts the fragmentation at the end of the 
splitting process for interaction strength $\lambda_0=0.02$ as a function of the radius $R$. 
The splitting rate is $\beta=1$.
The dynamical splitting process leads to fragmentation over the entire examined range of radii $R$.
For most of these radii the ground state of the system (at any barrier height) is condensed.
The system can thus dynamically fragment even though the ground state is condensed. 
Compared to the static phase diagram, Fig.~\ref{f2}b, 
the regime of dynamical fragmentation of the 2D BEC is significantly larger. 

Regardless of whether the ground-state density is located in the inner disk, 
the outer annulus, or in both parts of the trap, 
the fragmented final state is spread over the entire trap, 
see Fig.~\ref{f3}. This generic feature can be understood from the two opposing forces acting on the BEC. 
On the one hand,
the 2D attractive term $V_{2D}(r)$ tends to localize the particles in the inner disk. 
The repulsive interaction, on the other hand, naturally tends to push them apart from one another.
This competition promotes the dynamical spread of the BEC over 
the inner disk {\it and} the outer annulus in the splitting process.

To have a broader picture of the physical process 
we also study the fragmentation of the system as a function of the splitting rate $\beta$. 
Choosing a radius for which the ground state is condensed, $R=3$, 
we varied $\beta$ over two orders of magnitude, see Fig.~\ref{f4}b.
This cut through the dynamical phase diagram produces a broad region 
of splitting rates in which the system dynamically fragments. 
For slow rates ($\beta<0.1$) the system remains condensed throughout the splitting process. 
Interestingly, for fast rates ($\beta>10$) the system also remains condensed.
This suggests that the system requires a finite amount of time in order to fragment and 
pumping more energy into the system does not necessarily lead to larger fragmentation. 
In between, the dynamical splitting process leads to fragmentation 
over two orders of magnitude of $\beta$.

Fragmentation involves transferring of bosons out 
of the condensed mode.
This means that the condensed initial state must overlap
with a manifold of excited states with successively-increasing degree of fragmentation.
If the splitting process is too slow, 
the system does not reach these states,
whereas if it is too fast, 
there is no time to efficiently go through
such a manifold of fragmented excited states.

In conclusion, the present research investigates the many-body
physics of splitting a 2D BEC by a radial barrier.
We determine the static phase diagram which demonstrated the resilience of a 2D BEC to fragment. 
The ground state can only fragment in the vicinity of the degeneracy of GP energy of the two parts of the potential. 
The position of this degeneracy and the width of the fragmented region depend on the interaction strength. 
We then explore the dynamical process of splitting the BEC by a time-dependent barrier. 
This yielded a dynamic phase diagram which revealed that the 
system fragments over a much larger region compared to the static results. 
Strikingly, the dynamical fragmentation of a BEC, 
despite its ground state being fully condensed, 
was thus identified.
This opens up exciting possibilities beyond the current practice, 
that in order to produce in the splitting process a fragmented BEC,
the ground state must be fragmented. 
Furthermore,
our study suggests that a large manifold of fragmented excitations 
can significantly impact the dynamics of trapped 2D BECs.

As an outlook we mention that
implementing many-body linear response in 2D would provide the low-lying excitations which 
are not recovered by standard methods \cite{LR}. This would shed further light on 
the present findings.
We also speculate that the effect of dynamical fragmentation 
could be relevant in other circularly-shaped setups,
such as in Ref.~\cite{Nature_14}.
We believe the present work will stimulate the experimental and theoretical exploration 
of many-body dynamics in these systems.

\section*{Acknowledgements}

Financial support by the DFG is gratefully acknowledged. AUJL acknowledges financial support by the Swiss SNF and the
NCCR Quantum Science and Technology.
Computation time on the Cray XE6 system Hermit and
the NEC Nehalem cluster Laki at the HLRS, and the bwGRiD cluster
are gratefully acknowledged.




\begin{thebibliography}{99}
\bibitem{units1} To convert to dimensionless units 
the Hamiltonian is divided by $\frac{\hbar^2}{mL^2}$
where $m$ is a mass of a boson and $L$ a length scale.
As a concrete example we choose $L=2 \mu$m 
and $^{87}$Rb atom as a boson.
We point out that other realistic choices can be made.
The unit of energy $\frac{\hbar^2}{mL^2}$ corresponds to $29.1$Hz
and the unit of time $\frac{mL^2}{\hbar}$ to $5.47 \cdot 10^{-3}$sec.
The trapping potential height is $V=5.82$kHz and the
radius of the crater is $r_c=18 \mu$m.
\bibitem{units2} The relation between the dimensionless interaction parameter
$\lambda_0$, and the scattering length $a_s$
and the transverse confinement $l_z=\sqrt{\frac{\hbar}{m\omega_z}}$ along the $z$ axis 
is given by $\lambda_0 = 2 \sqrt{2\pi} \frac{a_s}{l_z}$,
where \cite{GSL} is used.
To realize the cases considered in this work, $a_s$ and $\omega_z$ can be chosen as follows.
Let the aspect ratio be $\frac{l_z}{2r_c} = \frac{1}{100}$, 
i.e., $\omega_z = 2\pi \cdot 897$Hz.
Then for $\lambda_0=0.002$ $a_s=0.144$nm,
for $\lambda_0=0.02$ $a_s=1.44$nm,
and for $\lambda_0=0.2$ $a_s=14.4$nm.
\end{thebibliography}
\end{document}